\documentclass[prc,preprint]{revtex4-1}
\usepackage{amssymb}
\usepackage[breaklinks]{hyperref}
\newcommand{\NPA}[1]{{\rm Nucl.~Phys.~A} {\bf #1}}
\newcommand{\NPB}[1]{{\rm Nucl.~Phys.~B} {\bf #1}}
\newcommand{\PLB}[1]{{\rm Phys.~Lett.~B} {\bf #1}}
\newcommand{\PRL}[1]{{\rm Phys.\ Rev.\ Lett.} {\bf #1}}
\newcommand{\PRD}[1]{{\rm Phys.~Rev.~D} {\bf #1}}
\newcommand{\PRC}[1]{{\rm Phys.~Rev.~C} {\bf #1}}
\newcommand{\RMP}[1]{{\rm Rev.\ Mod.\ Phys.} {\bf #1}}
\begin{document}
\title{Implications of the Oklo phenomenon\\
       in a chiral approach to nuclear matter} 
%       using relativistic nuclear structure theory}
%\titlerunning{Implications of the Oklo phenomenon}
\author{Edward D. Davis}
\email{edward.davis@ku.edu.kw}
\affiliation{Physics Department, North Carolina State University, Raleigh, North Carolina 27695-8202, USA\\
Department of Physics, Kuwait University, P.O.~Box 5969, 13060 Safat, Kuwait}
%           Fax: +965-2481-9374\\

\begin{abstract}
It has been customary to use data from the Oklo natural nuclear reactor  to place bounds on the change that 
has occurred in the 
electromagnetic fine structure constant $\alpha$ over the last 2 billion years. Alternatively, an analysis could 
be based on a recently proposed expression for shifts in resonance energies which relates them to changes in
both $\alpha$ and the average $m_q$ of the $u$ and $d$ current quark masses, and which makes explicit the
dependence on mass number $A$ and atomic number $Z$. (Recent model independent results on hadronic 
$\sigma$-terms suggest sensitivity to the strange quark mass is negligible.) The most sophisticated analysis, 
to date, of the quark 
mass term invokes a calculation of the nuclear mean-field within the Walecka model of quantum hadrodynamics. 
We comment on this study and consider an alternative in which the link to low-energy quantum chromodynamics 
(QCD) and its pattern of chiral symmetry-breaking is more readily discernible. 
Specifically, we investigate the sensitivity
to changes in the pion mass $M_\pi$ of a single nucleon potential determined by an \emph{in-medium\/} chiral 
perturbation theory ($\chi$PT) calculation which includes virtual $\mathrm{\Delta}$-excitations. 
Subject to some reasonable 
assumptions about low-energy constants (LECs), we confirm that the $m_q$-contribution to resonance shifts is enhanced
by a factor of 10 or so relative to the $\alpha$-term and deduce that
the Oklo data for Sm imply that $|m_q(\mathrm{Oklo})-
m_q(\mathrm{now})| \lesssim 10^{-9}m_q(\mathrm{now})$.
%
% Include up to five keywords.
\end{abstract}


\maketitle

\section{Introduction}\label{intro}

Dirac fathered some of the ideas central to light-front physics, the primary concern of this workshop. He 
also was the first physicist to speculate in print that the fundamental constants of nature may vary over 
cosmological time scales. His 650-word letter on the idea to Nature \cite{Dir37}, written within weeks 
of his honeymoon, was not all that well received:
``[a]s soon as Bohr finished reading the letter for the first time, he walked into Gamow's room in the 
Copenhagen institute and said, \emph{`Look what happens to people when they get married.'\/}\,'' (see 
p.~288 of Ref.~\cite{Far09}, my italics).
Today, however, there are many extensions of the Standard Model (SM) which 
naturally imply that fundamental parameters in the
SM Lagrangian are dynamical variables \cite{Uza2011}. There are also empirical indications \cite{WMFD2001} that 
the fine structure constant $\alpha$ may have changed over the lifetime of the universe (see the quasar 
absorption result in Table \ref{tb:bdsonalpha}).

The Oklo uranium ore mine in Gabon is the site of natural fission reactors that were active about
2 billion years ago. Sustained fission chain reactions occurred in seams of uranium ore about 1~m thick.
The record of this activity is to be found in the anomalous distribution of isotopes in the ores mined.
The isotopic abundance of ${}^{235}$U, for example, is 0.600\% as opposed to the natural abundance of 0.720\%.
The discovery of the existence of the Oklo reactors has prompted several intriguing lines of inquiry 
(see Refs.~\cite{Her2014} and \cite{DGS2014} for the most recent reviews), but of interest to us is the access 
it gives to compound nucleus reaction rates 2 billion years ago. Compound nucleus reactions can benefit
from enormous resonance enhancements, a fact which has been exploited in studies of parity and time-reversal 
non-invariance \cite{MDH1990,DH1991}. Such enhancements are also relevant to the Oklo 
data \cite{Shl76}.

The capture reaction $n\,+\,{}^{149}\mathrm{Sm}$ possesses a resonance near threshold, currently at a neutron 
energy of $E_r=97.3\,\mathrm{meV}$. Even a small change over time in the resonance energy $E_r$ would translate 
into a dramatic variation in the rate of capture by $^{149}$Sm of thermal neutrons. Conversely, the failure to
identify any difference between this capture rate now and at the time when the Oklo reactors were active would, 
in principle, imply a stringent
bound on the shift in resonance energy $\Delta E_r \equiv E_r(\mathrm{Oklo}) - E_r(\mathrm{now})$. In turn, one 
could infer non-trivial bounds on the change with time of parameters in the nuclear Hamiltonian, since $E_r$ is
related to its eigenenergies.
In practice, the bound on $\Delta E_r$ is circumscribed by uncertainties in the modeling of the operation of the Oklo 
reactors, and the extraction of the corresponding limits on the time dependence of Hamiltonian parameters
is frustrated by the complexities of the nuclear many-body problem. Nevertheless, as Table 
\ref{tb:bdsonalpha} shows, analysis of Oklo data has yielded the \emph{most restrictive bound\/} to date on the 
variation of the fine structure constant $\alpha$ with redshift $z$.

\begin{table*}[h]
\caption{Bounds on $\alpha(z)-\alpha_\mathrm{now}$ 
(adapted from Ref.~\cite{Chi011} with the Oklo result taken from Ref.~\cite{Oneg012})}
\centering
\label{tb:bdsonalpha}
\begin{tabular}{lccc}
\hline\noalign{\smallskip}
                           &$z$& $[\alpha(z)-\alpha_\mathrm{now}]/\alpha_\mathrm{now}$ 
                                                               & $\dot{\alpha}/\alpha\ (\mathrm{yr}^{-1})$
 \\ \hline
%\tableheadseprule\noalign{\smallskip}
  Atomic clock (Al$^+$/Hg$^+$)
                               &     0   &                     &   $(-1.6\pm 2.3)\times 10^{-17}$      \\
  Oklo ($\mathrm{n}+{}^{149}\mathrm{Sm}$)
                               &    0.16 & $(-1.0\mapsto 0.7)\times 10^{-8}$ & $(-4\mapsto 5)\times 10^{-18}$     \\
  Meteorites                   &    0.43 & $(-0.25\pm 1.6)\times 10^{-6}$    &                                    \\
  Quasar absorption (MM)       &$0.2-4.2$& $(-5.7\pm 1.1)\times 10^{-6}$     &                                    \\
  Cosmic $\mu$wave background  &  $10^3$ &  $-0.013\mapsto 0.015$            &                                    \\
  Big-bang nucleosynthesis     &  $10^9$ &  $< 6\times 10^{-2}$              &                                    \\ 
\noalign{\smallskip}\hline
\end{tabular}
\end{table*}

In the next section, we discuss a recent proposal \cite{DGS2014} for the interpretation of Oklo data 
which accommodates changes in \emph{both\/} the fine structure constant $\alpha$ and $X_q=m_q/\Lambda$, 
where $\Lambda$ is the mass scale of QCD. 
One undecided issue is the relative magnitude of the two 
contributions. We point out relevant features of earlier estimates~\cite{Flam09} of the sensitivity to 
changes in $X_q$ and then, in section \ref{sc:Interpretation}, present an attempt to corroborate this work using 
a chiral effective field theory ($\chi$EFT) model for symmetric nuclear matter \cite{KFW2002a,KFW2002b,FKW2005}. 
Conclusions are drawn in section \ref{sc:concl}.

\section{Towards a unified treatment of Oklo data on resonance shifts}

A synthesis \cite{DGS2014} of the results in Refs.~\cite{Flam09} and \cite{Dam96} implies that shifts
for neutron capture resonances
\begin{equation}\label{eq:Oklo}
   \Delta E_r = a \frac{\Delta X_q}{X_q} + 
                           b \frac{Z^2}{A^\frac{4}{3}}\frac{\Delta\alpha}{\alpha} ,
\end{equation}
where, significantly, it is conjectured that the coefficients $a$ and $b$ are approximately \emph{independent\/} 
of the mass number $A$ and the atomic number $Z$ of the target. 
The considerations of Ref.~\cite{DGS2014} about the magnitudes of $a$ and $b$ may be summarized as follows: to 
within a factor of 2 or so, $|b| \sim 0.5\,\mathrm{MeV}$, but the order of magnitude of $a$ is uncertain.

On the basis of pain-staking variational Monte Carlo calculations for p-shell nuclei, using the 
sophisticated Argonne $v_{18}$ plus Urbana IX interactions, the authors of Ref.~\cite{Flam09} claim that 
$a\sim 10\,\mathrm{MeV}$. However, their result is extremely sensitive to the properties of an exchange boson
(of mass $m_V$), which is introduced to mock up the short-range repulsion associated with heavy mesons. The
crucial parameter is the dimensionless sensitivity coefficient
\[
 K_V^q \equiv  \frac{X_q}{m_V}  \frac{\delta m_V}{\delta X_q}
\]
for which there is no first principles determination. Alternative choices of the value of $K_V^q$,
which cannot be ruled out with the information at our disposal, change the value of $a$ by \emph{a factor of 10
or more} --- see section 6.2 of Ref.~\cite{DGS2014} for more details (the values of  $K_V^q$ used to generate 
sets 2a, 2b and 2c of Table 4 in Ref.~\cite{DGS2014} are 0.06, 0.07 and 0.08, respectively, not 0.6, 0.7 and 0.8 
as stated in the caption of this Table and elsewhere in the text).

%
% Work of Olive?
%

Within a schematic treatment of compound nucleus states, which presupposes that the nucleons are moving in an
attractive square well potential of depth $U_0$ and radius $R=r_0 A^\frac{1}{3}$,
the shift in $E_r$ due to a change $\delta X_q$ in $X_q$ is given by \cite{DmFl2003}
\begin{equation}\label{eq:DE}
 \delta E_r \approx -U_0 \left( 
                                \frac{\delta m_N}{m_N} + 2\frac{\delta r_0}{r_0} + \frac{\delta U_0}{U_0}
                         \right),
\end{equation}
where $\delta m_N$, $\delta r_0$ and $\delta U_0$ denote the related changes in the nucleon mass $m_N$, the radius 
parameter $r_0$ and the potential well depth $U_0$, respectively. Equation (\ref{eq:DE}), which should be adequate 
for order of magnitude estimates, is the starting point for another estimate of $a$ in Ref.~\cite{Flam09}, this
time using the Walecka model. The scalar ($S$) and vector ($\cal V$) boson-nucleon couplings advocated in 
Ref.~\cite{BrWe1977} are adopted and all terms involving $\delta r_0$ are discarded to yield
\begin{eqnarray}\label{eq:Walest}
 \delta E_r &\approx & U_0 \left( 7.50\frac{\delta m_{ S}}{m_{ S}} 
                           - 5.50 \frac{\delta m_{\cal V}}{m_{\cal V}}-\frac{\delta m_N}{m_N} \right) \nonumber\\
            &= &    U_0 \left( 7.50 K_{ S}^q - 5.50 K_{\cal V}^q - K_N^q \right) \frac{\delta X_q}{X_q}.
\end{eqnarray}
Unfortunately, there seems to be an unavoidable element of arbitrariness in the assignment of values to the 
coefficients $K_{\cal V}^q$ and, in particular, $K_{ S}^q$. 

The authors of Ref.~\cite{Flam09} are comfortable about identifying the vector meson with the $\omega$ meson, but 
admit that the scalar meson ``imitates both the $\sigma$ meson exchange and two-pion exchange''.
This characterization of the scalar meson is debatable in as much as many regard the $\sigma$-meson as a
fictitious artifact of phenomenological one-boson exchange models. It has been known for some time \cite{KGW98} 
that a parameter-free isoscalar central potential generated by 2$\pi$-exchange with single and double 
$\mathrm{\Delta}$-excitation is in excellent agreement with the $\sigma$-exchange potential for distances 
$r > 2\,\mathrm{fm}$. Another complication is that no model-independent estimate of $K_\omega^q$ exists.
Of more concern is the suggestion \cite{FKVW2004} that the strong scalar and vector mean fields of the Walecka 
model have nothing whatsoever to do with exchange bosons, but are, instead, induced by changes of QCD 
vacuum condensates in the presence of baryonic matter. A scheme marrying this idea with chiral 
$\pi\mathrm{N}\mathrm{\Delta}$ dynamics
of the kind considered in the next section is found to work very well for a broad range of spherical
and deformed nuclei \cite{FKVW2006}. 
 
In the remainder of this paper, we focus on an alternative estimate of the $\delta U_0$-term in Eq.~(\ref{eq:DE}).
It is guided by the observation that, in the studies of Refs.~\cite{FKVW2004} and \cite{FKVW2006},
binding (and, hence, $U_0$) is predominantly accounted for by the chiral dynamics.

\section{Interpretation of Oklo within a $\chi$EFT for nuclear matter}\label{sc:Interpretation}

Chiral effective field theories provide a framework for the development of a sounder basis for the theory of nuclear
forces \cite{EHM09}. Of specific interest to us are the opportunities  $\chi$EFTs offer for the investigation of
dependence on the pion mass $M_\pi$ or, via the Gell-Mann-Oakes-Renner relation, the average of the light
quark masses. The relation of the properties of two-nucleon systems to $M_\pi$ was first
discussed within the context of $\chi$EFTs a little over a decade ago \cite{Bean2003a,Bean2003b,EMG03}.
%
% Combine these references into a single citation 
% 
Quite recently, this type of analysis has been extended to the light nuclei pertinent to Big Bang nucleosynthesis 
\cite{BLP2011,BEFH2013}, and the computationally daunting problem of the Hoyle state in ${}^{12}$C \cite{EKLL2013}. 
Such detailed treatments are not feasible for a heavy nucleus like ${}^{150}$Sm, but the chiral approach to
nuclear matter developed in Refs. \cite{KFW2002a} and \cite{FKW2005} will suffice for an 
order of magnitude estimate.
(In this section, units are chosen so that $\Lambda_{QCD}=1$, meaning that $X_q=m_q$.)
%
% Reintroduce because of appearance of (delta M_pi)/M_pi in several equations
%

\subsection{$\chi$EFT model for the real part of the single particle potential in nuclear matter}  

The model of Ref.~\cite{KFW2002a} (as extended in Ref.~\cite{FKW2005}) proceeds from the the recognition that, 
at or near the saturation density of symmetric nuclear matter, the magnitude of the Fermi momentum $k_F$, 
the pion mass $M_\pi$ and the $\Delta\!-\!\mathrm{N}$ mass difference are all comparable and small relative to 
the chiral symmetry-breaking scale $\Lambda_\chi$. Thus, in considering interactions of long and 
intermediate range, pions are taken to be explicit degrees of freedom and the effect of virtual 
$\mathrm{\Delta}$-excitations are incorporated. More precisely, the corresponding 
$\pi\mathrm{N}\mathrm{\Delta}$ dynamics are treated using in-medium 
$\chi$PT \cite{KFW2002a}. The calculations \cite{KFW2002b,FKW2005} of the real part $U$ of the associated single 
particle potential sum the following contributions (we adopt the classification scheme of section 3 in 
Ref.~\cite{KFW2002b}): 
(a) 1$\pi$-exchange Fock diagrams with two medium insertions;
(b) twice-iterated 1$\pi$-exchange Hartree diagrams with two and three medium insertions;
(c) twice-iterated 1$\pi$-exchange Fock diagrams with two and three medium insertions;
(d) irreducible 2$\pi$-exchange Fock diagrams with two medium insertions and 0, 1 or 2 intermediate 
    $\mathrm{\Delta}$-excitations;
(e) Hartree diagrams with three medium insertions and a single intermediate $\mathrm{\Delta}$-excitation, and;
(f) Fock diagrams with three medium insertions and a single intermediate $\mathrm{\Delta}$-excitation.

Ultraviolet divergent pion-loop diagrams in (d) are regularized by using suitably subtracted spectral
representations. Within this model, the two subtraction constants ($B_3$ and $B_5$, in the notation of 
Ref.~\cite{FKW2005}) are interpreted as LECs related to 2-body contact interactions, accommodating the 
unresolved short-range dynamics. In fact,
$B_3$ subsumes as well linear divergences arising from diagrams in (c) containing two medium insertions.
It is also found helpful to introduce a third LEC, $\zeta$, which determines the strength of a three-body contact
interaction designed to eliminate the quadratic dependence on nuclear density of the Hartree diagrams in (e).
(This quadratic dependence is inconsistent with the known saturation properties of nuclear matter.)

In a departure from the standard treatment of LECs in $\chi$EFTs, the values of these constants
are chosen so that semi-empirical saturation properties of nuclear matter (i.e.~binding energy per particle, 
density and compressibility) are adequately reproduced (for details, see the penultimate page of section 2
in Ref.~\cite{FKW2005}). Despite the expedient nature of this procedure and the ad hoc character of the 3-body contact
interaction, the model is phenomenologically satisfactory and in good agreement with sophisticated many-body 
analyses (e.g., Dirac-Brueckner calculations based on realistic NN-potentials). The authors of Ref.~\cite{FKW2005} 
suggest that their approach to fixing LECs may be justified because the equivalent contact interactions could
``represent the full content of the short distance T-matrix'' and ``should therefore not be iterated with 
long-range pion-exchange pieces'' (in contradistinction to earlier work~\cite{LFA2000}).

\subsection{Sensitivity of $U_0$ to $m_q$: contributions of long- and intermediate-range interactions}

The strength $U_0$ of the real part of the single particle potential is found from the results
in Refs.~\cite{KFW2002b} and \cite{FKW2005} by taking the limit in which the magnitude of the single particle 
momentum $p\rightarrow 0$ (sometimes with the aid of L'H\^{o}pital's rule). The part of $U_0$ arising from 
long- and intermediate-range interactions is a function of $k_F$ and five hadronic parameters: the pion mass 
$M_\pi$, the pion decay constant $F_\pi$, the nucleon axial coupling constant $g_A$, the nucleon mass $m_N$ 
and the delta-nucleon mass splitting $\Delta$. For the sake of illustration, we quote the contribution to $U_0$ 
deriving from the diagrams with \emph{two} medium insertions listed under (b) above (and, hence, denoted as
$U_{0b(2)}$):
\[
  \frac{U_{0b(2)}}{m_N} = \frac{\pi}{4}\left(\frac{g_A M_\pi}{2\pi F_\pi}\right)^4 
                                                                    \left[(9+6 u^2)\tan^{-1}u - 9u \right],
\]
where $u=k_F/M_\pi$. Actually, $U_{0b(2)}$ has the distinction of being larger (near the saturation point of 
symmetric nuclear matter) than any of the other pieces of $U_0$ computed from the diagrams in (a)-to-(f) 
above (in our notation, $U_{0a}$, $U_{0b(3)}$, $U_{0c(2)}$, etc). The fractional sizes $f_i\equiv U_{0i}/U_0$ of the 
$U_{0i}$'s are given in Table~\ref{tb:dU} for the Fermi momentum $k_{F0}$ corresponding to the 
saturation density of symmetric nuclear matter. Here, and in the remainder of this paper, we adopt the values 
used in Ref.~\cite{FKW2005} for $k_{F0}$ and the five hadronic parameters listed above.

\begin{table*}[h]
\caption{Fractional sizes $f_i\equiv U_{0i}/U_0$ and related sensitivity coefficients $K_i^\pi$}
\centering
\label{tb:dU}
\begin{tabular}{*{9}{c}}
\hline\noalign{\smallskip}
$i$      &    a   &              b(2) &    b(3) &              c(2) &                c(3)  &             d   &   e  
                          &             f       \\ \hline
%\tableheadseprule\noalign{\smallskip}
$f_i$ 
         &$-0.364$&            $-1.34$& $-0.995$&           $-0.743$&\hphantom{$-$}$0.0358$&       $-0.0458$ & 
$-0.346$          & \hphantom{$-$}0.125 \\ 
$K_i^\pi$&$-0.639$&\hphantom{$-$}1.38 & $-1.09$\hphantom{5}
                                               &\hphantom{$-$}0.283& $-1.91$\hphantom{58} &$-0.748$\hphantom{8}& 
          $-1.03$\hphantom{2}&  $-1.08$\hphantom{5}\\ 
\noalign{\smallskip}\hline
\end{tabular}
\end{table*}

In terms of the hadronic parameters $P=\{M_\pi,F_\pi,g_A,m_N,\Delta\}$, the change in $\tilde{U}_0=\sum_i U_{0i}$
induced by a change in $m_q$ is
\begin{equation}\label{eq:dU}
  \delta\tilde{U}_0
  = \Biggl[ 
    \sum\limits_{P,\, i}
                 U_{0i} \left( \frac{P}{U_{0i}} \frac{\delta U_{0i}}{\delta P} \right)
                  \left( \frac{m_q}{P} \frac{\delta P}{\delta m_q} \right) 
    \Biggr] \frac{\delta m_q}{m_q} ,
\end{equation}
where the factorization into sensitivity coefficients 
\[
K_{U_{0i}}^P \equiv \frac{P}{U_{0i}} \frac{\delta U_{0i}}{\delta P}
\qquad\mbox{and}\qquad 
K_P^q \equiv \frac{m_q}{P} \frac{\delta P}{\delta m_q}
\]
proves convenient in Eq.~(\ref{eq:dUtilde}) below and 
permits us to invoke existing results on the $K_P^q$'s. \emph{Unlike the sensitivity coefficients $K_{S}^q$ 
and $K_{\cal V}^q$ of Eq.~(\ref{eq:Walest}), all the $K_P^q$'s can be, in principle, unambiguously determined.} 
The careful assessment in Ref.~\cite{BEFH2013} 
of the best estimates for the $K_P^q$'s indicates that $K_{M_\pi}^q = 0.49$ (to two significant figures) 
and that all other $K_P^q$'s are at least an order of magnitude smaller. Consequently, we retain in 
Eq.~(\ref{eq:dU}) only the $M_\pi$-term. We also employ the notation $K_i^\pi$ instead of the more fastidious
$K_{U_{0i}}^{M_\pi}$.

Calculation of the sensitivity coefficients $K_i^\pi$ is straightforward. The results are presented in 
Table \ref{tb:dU}. In terms of the $K_i^\pi$'s, the fractional change $\delta\tilde{U}_0/U_0$
implied by a change $\delta M_\pi$ in $M_\pi$ is
\begin{equation}\label{eq:dUtilde}
  \frac{\delta\tilde{U}_0}{U_0} =  \Bigl[ \sum\limits_i f_i K_i^\pi \Bigr] \frac{\delta M_\pi}{M_\pi},
\end{equation}
which, using the results in Table \ref{tb:dU} and the value of $K_{M_\pi}^q$ given in the previous paragraph, 
reduces to
\begin{equation}\label{eq:dtU}
 \frac{\delta\tilde{U}_0}{U_0} =  -0.56\frac{\delta M_\pi}{M_\pi} 
                               =  -0.56 K_{M_\pi}^q \frac{\delta m_q}{m_q} = -0.28 \frac{\delta m_q}{m_q} .
\end{equation}
If we take $K_{m_N}^q=0.048$ (from Ref.~\cite{BEFH2013}), $U_0\simeq 50\,\mathrm{MeV}$, and discard 
the $r_0$-term in Eq.~(\ref{eq:DE})  (as the authors of Ref.~\cite{Flam09}, in effect, do), then 
Eqs.~(\ref{eq:DE}) and (\ref{eq:dtU}) imply the order of magnitude estimate $a\sim 10\,\mathrm{MeV}$, which is 
\emph{the same as the estimate\/} of Ref.~\cite{Flam09}. However, we have still to include the contact 
interactions in our analysis.

\subsection{Sensitivity of $U_0$ to $m_q$: effect of contact interactions}

In full, the part of $U_0$ related to contact interactions is
\begin{equation}\label{eq:contact}
  \breve{U}_0 = 2 \frac{B_3}{m_N^2} k_F^3 +  \frac{B_5}{m_N^4} k_F^5
                                          + 2 \frac{\zeta}{\Delta}  \left(\frac{ g_A}{2\pi F_\pi} \right)^4 k_F^6 .
\end{equation}
For $k_F = k_{F0}$ and the preferred choice of ($B_3,B_5,\zeta$) made in Ref.~\cite{FKW2005} 
(i.e., $B_3=-7.99$, $B_5=0$ and $\zeta=-{\textstyle\frac{3}{4}}$), the first term in Eq.~(\ref{eq:contact}) is an 
order of magnitude larger than the remaining terms. Retaining only the $B_3$-term and paralleling the analysis
leading to Eq.~(\ref{eq:dUtilde}), we find that
\[
   \frac{\delta \breve{U}_0}{U_0} \sim 4.15 K_{B_3}^{M_\pi} \frac{\delta M_\pi}{M_\pi} .
\]
We can arrive at an estimate of $K_{B_3}^{M_\pi}$ by taking advantage of the following approximate 
relation between $U_0$ and momentum-space matrix elements of the universal low-momentum NN-potential
$V_{\mathrm{low}\; k}$ \cite{BFS2010}:
\begin{equation}\label{eq:lowV}
      U_0 = \frac{3\pi}{2m_N}\Bigl[
                                  V^{({}^1S_0)}_{\mathrm{low}\; k}(0,0) + V^{({}^3S_1)}_{\mathrm{low}\; k}(0,0) 
                                  \Bigr]  \rho,
\end{equation}
which applies
in the limit of vanishing nuclear density $\rho$~\cite{FKW2005}. The part of $U_0$ linear in $\rho$ reads,
quite generally,
\begin{equation}\label{eq:linearU}
     \frac{3\pi}{2m_N} \left[ \frac{2\pi}{m_N} B_3
                              + {\textstyle\frac{15}{16}}\pi^2\left(\frac{g_A}{2\pi F_\pi}\right)^4 m_N^2 M_\pi 
                      \right] \rho ,
\end{equation}
where the the term containing $M_\pi$ originates from $U_{0b(2)}$ and $U_{0c(2)}$. Under the assumption that
the $M_\pi$-dependence of matrix elements of $V_{\mathrm{low}\; k}$ is negligible (in view of the manner of its
construction), Eqs.~(\ref{eq:lowV}) and (\ref{eq:linearU}) imply that
\[
   K_{B_3}^{M_\pi} \approx -\frac{15}{32}\frac{\pi}{B_3} 
                            \left(\frac{g_A m_N}{2\pi F_\pi}\right)^4 \frac{M_\pi}{m_N} = 0.52,
\]
which, in turn, implies that $\delta \breve{U}_0\left/ U_0\right. \sim 1.1 \delta m_q/m_q$. Combining this last
result with that in Eq.~(\ref{eq:dtU}), we deduce that $\delta U_0/U_0 \sim 0.8 \delta m_q/m_q$. Our final estimate
of $a$ is, accordingly, $a\sim -40\,\mathrm{MeV}$.

\section{Conclusion}  \label{sc:concl}

% Begin with a reference to Langacker's hope  (No space)

We have attempted to put the determination with Oklo data of $\Delta X_q=X_q(\mathrm{Oklo})-X_q(\mathrm{now})$
on a firmer theoretical footing. Our analysis confirms earlier claims that Oklo data are more 
sensitive to $\Delta X_q$ than to $\Delta\alpha$: referring to Eq.~(\ref{eq:Oklo}), our estimate for the 
coefficient of $\Delta X_q/X_q$ is more than a factor of 10 bigger than the value of the coefficient 
of $\Delta\alpha/\alpha$ for ${}^{149}$Sm ($Z^2/A^\frac{4}{3}=4.87$ for $^{149}$Sm). The effect of the 
$r_0$-term in Eq.~(\ref{eq:DE}) (also ignored in earlier work) has still to 
be established, but, supposing it to be negligible, the bound of $|\Delta E_r|<11\,\mathrm{meV}$~\cite{Oneg012} 
for the $^{150}$Sm resonance, coupled with our estimate of $|a|$, which we conservatively take to be
$|a|\sim 10\,\mathrm{MeV}$, implies the bound
% ($\sim 40\,\mathrm{MeV}$), implies the bound
$|\Delta m_q| \lesssim 1\times 10^{-9} m_q(\mathrm{now})$.

\begin{acknowledgements}
I would like to thank the organizers of LightCone 2014 for giving me the opportunity to present my work.
% and for creating such a pleasant ambiance. 
I also thank Prof.~N.~Kaiser for his efforts in
responding to my queries and the Physics Department at NCSU for its hospitality. 
%Finally, I want to 
%acknowledge the debt I owe to Chris Gould for introducing me to the Oklo phenomenon and for his more 
%than generous subsequent encouragement. 
\end{acknowledgements}


%\bibliographystyle{spphys}  
%\bibliography{oklo}


\end{document}